\documentclass[twocolumn,preprintnumbers,superscriptaddress,endnote,nofootinbib,prl]{revtex4}
\usepackage{graphicx}

\usepackage[hypertex]{hyperref}
\newcommand{\vev}[1]{\langle {#1} \rangle}
\newcommand{\lsim}{\lesssim}
\newcommand{\gsim}{\gtrsim}

\newcommand{\eq}[1]{Eq.~(\ref{#1})}

\newcommand{\ord}[1]{\mathcal{O}{(#1)}}
\newcommand{\beq}{\begin{equation}}
\newcommand{\eeq}{\end{equation}}
\newcommand{\eps}{\varepsilon}

\newcommand{\gamd}{\gamma_d}
\newcommand{\ald}{\alpha_d}
\newcommand{\mgd}{m_{\gamma_d}}

\newcommand{\mP}{M_{\rm P}}

\begin{document}

\pagestyle{plain}

\title{\boldmath Running of the $U(1)$ coupling in the dark sector}

\author{Hooman Davoudiasl
\footnote{email: hooman@bnl.gov}
}
\author{William J. Marciano
\footnote{email: marciano@bnl.gov}
}

\affiliation{Department of Physics, Brookhaven National Laboratory,
Upton, NY 11973, USA}


\begin{abstract}

The ``dark photon" $\gamma_d$ of a gauged $U(1)_d$ can
become practically invisible if it primarily decays into light states from a dark sector.
We point out that, in such scenarios, the running of the
$U(1)_d$ ``fine structure constant" $\ald$,
with momentum transfer $q^2$, can be significant and potentially measurable.
The $\gamma_d$ kinetic mixing parameter $\varepsilon^2$ is also expected
to run with $q^2$, through its dependence on $\alpha_d$.
We show how the combined running of $\varepsilon^2 \alpha_d$ may provide a probe of the spectrum of dark particles
and, for $\ald\gsim {\rm few}\times 0.1$, substantially modify
predictions for ``beam dump" or other intense source experiments.  These features are demonstrated in simple
models that contain light dark matter and a scalar that breaks $U(1)_d$.  We also discuss
theoretic considerations, regarding the $U(1)_d$ model in the ultraviolet regime,
that may suggest the infrared upper bound $\ald \lsim 0.1$.

\end{abstract}
\maketitle

In recent years, various phenomenological considerations have motivated
the introduction of a light vector boson, the ``dark photon" $\gamd$,  with a mass $\mgd \lsim 1$~GeV -
associated with a spontaneously broken gauged $U(1)_d$.  The dark photon couples
to the Standard Model (SM) only through kinetic mixing \cite{Holdom:1985ag} of dark charge
with hypercharge, parameterized
by $\eps\ll 1$.  At low energies,
$\gamd$ kinetically mixes with the photon and couples
primarily to the SM electromagnetic current with a strength
$\eps e$, where $e$ is the electromagnetic coupling.  A great
deal of theoretical and experimental effort has been directed towards
dark photon physics \cite{Bjorken:2009mm,Essig:2013lka}.  Important motivation for these efforts
have been provided by potential astrophysical signals of
dark matter \cite{ArkaniHamed:2008qn}, as well as the
$3.6\sigma$ deviation from the SM prediction of the
muon anomalous magnetic dipole moment $g_\mu-2$ \cite{Bennett:2006fi}
which can be explained by a light $\gamd$ and $\eps \approx 2\times 10^{-3}$ \cite{Pospelov:2008zw}.
Such small values of $\eps$ most naturally arise from loop effects \cite{Holdom:1985ag} of
particles charged under both the dark and SM hypercharge $U(1)$ interactions.  For example,
a typical 1-loop value might be $\eps \sim e g_d/(16 \pi^2)$.

If $\gamd$ is the lightest state in the dark sector then its decay
is mainly to charged SM states - typically leptons $\ell=e,\mu$ - which
could be used to detect it.  However, in the presence of light dark states,
in particular dark matter, $\gamd$ would decay mainly into those particles,
since in typical scenarios $g_d\gg \eps e$, where
$g_d$ is the coupling constant of the $U(1)_d$ gauge interactions.  The
different phenomenology of such a nearly ``invisible" $\gamd$ provides new possibilities
to explain various anomalies, such as the aforementioned $g_\mu-2$,
but is subject to different sets of experimental
constraints \cite{Izaguirre:2013uxa,Diamond:2013oda,Davoudiasl:2014kua,Batell:2014mga}.
In addition, the invisible $\gamd$ scenario offers an interesting opportunity for
producing and detecting sub-GeV dark matter in accelerator based experiments.
Boosted dark photons - produced in high intensity fixed target
experiments - decay in flight and lead to a ``dark matter beam" which can be detected downstream
\cite{Batell:2009di,deNiverville:2011it,Dharmapalan:2012xp,Izaguirre:2013uxa,Diamond:2013oda}.
The detection rate depends on $\gamd$ couplings to the dark and visible
sectors, $g_d$ and $\eps e$, respectively; see also Ref.~\cite{Essig:2013vha}.  The signal will look like
electron scattering, modulo radiative corrections and scaled by $\eps^2 g_d^2/e^2$.

The coupling of light dark matter to the dark photon can provide a
mechanism for obtaining its observed abundance as a thermal relic density.  Here, the annihilation
of dark matter via $\gamd$ into SM states is mediated by kinetic mixing which allows
one to derive a rough relation among the model
parameters. Using the results of Refs.~\cite{deNiverville:2011it,Izaguirre:2014bca},
one finds
\beq
\ald \sim 0.02\, w \left(\frac{10^{-3}}{\eps}\right)^2 \left(\frac{\mgd}{100~\text{MeV}}\right)^4
\left(\frac{10~\text{MeV}}{m_d}\right)^2\,,
\label{ald}
\eeq
where $\ald\equiv g_d^2/(4 \pi)$, $m_d$ is the mass of a dark matter state,
$w\sim 10$ for a complex scalar \cite{deNiverville:2011it},
and $w\sim 1$ for a fermion \cite{Izaguirre:2014bca}.  Here, and elsewhere
in this work, $\alpha_d$ and $\eps$ are defined by their values at low momentum
transfer $q^2 \sim \mgd^2$.
Given that some of the existing bounds \cite{Batell:2014mga} and proposed
experiments \cite{Izaguirre:2014bca} probe values of $\eps$ as low as $10^{-4}$,
\eq{ald} would then require $\ald\gsim 1$,
keeping other parameters at their above reference values, to get the
correct dark matter relic abundance.  Note that if dark matter density is set by an asymmetry,
efficient annihilation of its symmetric population would require somewhat larger
annihilation cross sections, compared to the $\sim$~pb implied by \eq{ald},
and hence even larger $\ald$ values.  Thus, generally speaking,
values of $\ald \gsim 0.01-0.1$ can be motivated if the dark photon
is assumed to decay primarily into dark matter states.

In this work, we point out that the
running of $\ald$  - as a function of momentum transfer $q$, due to quantum loops of
light dark sector states - can have important phenomenological
implications.\footnote{See, {\it e.g.}, Refs. \cite{Zhang:2009dd,Sannino:2014lxa} for possible cosmological
effects of running dark sector couplings, within different frameworks.}
In dark photon models it is generally assumed that the kinetic mixing parameter $\eps$
corresponds to $q^2 \approx 0$ and remains constant with increasing $q$.  However, as mentioned earlier,
kinetic mixing is naturally loop-induced and hence $\eps^2 \propto \alpha \alpha_d$,
with $\alpha\equiv e^2/(4\pi)$.  Thus, the running of $\alpha_d$, due to the
effect of light dark states, induces a similar running in $\eps^2$ that would lead to its growth
with increasing $q$ and the combination $\ald\, \eps^2$ actually grows like $\ald^2$.

\begin{figure}[t]
\includegraphics[width=0.46\textwidth,clip]{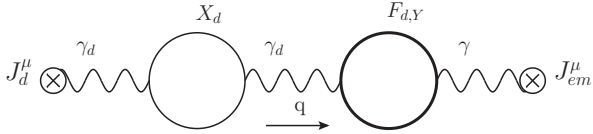}
\caption{Vacuum polarization correction to kinetic mixing.  The $U(1)_d$ current $J^\mu_d$ couples to the SM electromagnetic current
$J^\mu_{em}$, via kinetic mixing, mediated by quantum effects of heavy particles $F_{d,Y}$ (thick loop) that carry both
dark charge and hypercharge.  The vacuum polarization correction from light dark states $X_d$ (thin loop) leads to the
running of $\alpha_d(q)$ with momentum transfer $q$ which induces a similar running
in the kinetic mixing parameter $\eps^2(q)$}
\label{eps-run}
\end{figure}

The above effect is illustrated in Fig.\ref{eps-run}, where the vacuum polarization correction to
the interaction of a $U(1)_d$ current $J^\mu_d$ with the SM electromagnetic current $J^\mu_{em}$,
mediated by kinetic mixing, is illustrated.  In Fig.\ref{eps-run}, the thick loop, denoted by $F_{d,Y}$, represents
heavy states - charged under both the SM hypercharge and $U(1)_d$ - that
generate kinetic mixing at the quantum level.  The thin loop, denoted by $X_d$,
represents light dark states whose vacuum polarization contributions
lead to the running of $\ald(q)$.  The
loop-induced $\eps^2(q)\propto \alpha(q) \alpha_d(q)$ inherits its running from $\alpha_d(q)$.
The well-known running of $\alpha(q)$ due to vacuum polarization effects
in quantum electrodynamics is relatively small and, hence, ignored in
our discussions.

Given the discussion following \eq{ald},
values of $\ald$ not far below unity are well motivated if light dark matter communicates
with the SM through a heavier invisible $\gamd$ mediator.
We show that the growth of $\ald$ with momentum transfer $q$ can
constrain the regime of validity of calculations, depending on the choice of model parameters, and
may potentially lead to significant observable effects.

Although the existence of a Landau pole signals strong coupling behavior,
the onset of strong coupling, where perturbation theory starts to break
down, is probably closer to $\alpha_d\sim 1$.
From a theoretical point of view, the approach of a Landau pole for $\ald$ can be interpreted
as the onset of a new non-Abelian interaction, or some other ultraviolet completion, that would supplant the
low energy $U(1)_d$ and avoid an ever-growing interaction strength with larger $q$.
Such considerations can have interesting implications
for the underlying model at energy scales well above $\mgd$ \cite{Zhang:2009dd,Sannino:2014lxa},
but will not be studied here, except to mention their use as a potential constraint on $\alpha_d$.  

Before going further, we would like to remark that light vector
models with direct coupling to the SM have also been invoked in various contexts \cite{directU1}.
Such direct couplings have to be tiny, or else one would observe severe deviations from standard physics.
Much of the phenomenological discussion in our work would also apply to these models,
as long as one assumes $\ord{1}$ values for $\ald$ and dark sector charges $Q^{\rm DS}_d \sim 1$,
but small charges for SM fields $Q^{\rm SM}_d \lsim 10^{-3}$,
under the new Abelian gauge interactions \cite{Fayet:2004bw}.  We
will assume the alternative kinetic mixing picture in our work,
as it naturally yields suppressed interactions between the visible and the dark sectors (in principle,
$\gamd$ could also have mass mixing with the SM $Z$, leading to additional phenomenology \cite{Davoudiasl:2012ag}).

As our basic model, we will assume that the dark sector contains
a dark matter state, a fermion $\psi$ or a scalar $\phi$, as well as a dark Higgs particle
with non-zero vacuum expectation value $\vev{\Phi_d}$
that is responsible for the breaking of $U(1)_d$; for simplicity all these particles
are assumed to have unit charges $|Q_d|=1$ under $U(1)_d$.  We will focus
on the regime of momentum transfer $q > \mgd$, where symmetry breaking effects
are negligible and the running of $\ald$ is significant.  In typical
proposed fixed target experiments, 10~MeV $\lsim q \lsim$ GeV, where
$\mgd\gsim 10$~MeV is probed.  As we are focusing
on ``invisible" dark photon models, the dark matter state
($\psi$ or $\phi$) will be assumed lighter than $\gamd$.
Assuming that $\Phi_d$ is not strongly self coupled, it is quite natural
to expect that its mass parameter $m_{\Phi_d}\sim \mgd$ and hence typically less than $q$.
Hence, we will include both the dark matter state
and $\Phi_d$ contributions to the running of $\ald$ until $q\lsim \mgd$.  The infrared
value of the $U(1)_d$ coupling can thus be defined by its value at $q^2 = \mgd^2$, denoted by
$\alpha_d(\mgd)$.

Since we are focused on the kinematic
regime where $q> \mgd$, we will ignore
the mass of the vector boson, capturing the leading behavior as a function of $\mgd/q$.
This suffices for the purposes of our discussions and to highlight
the key features of generic invisible $\gamd$ scenarios.  Detailed calculations
for specific experiments and model parameters lie outside the scope of this work. 

In determining the regime
where $\ald \gsim 1$, higher order effects can become important and
we will therefore perform a 2-loop analysis.
The 2-loop beta function of $U(1)_d$, with $n_F$ fermions
and $n_S$ scalars of unit charge, is given by
(see, for example, Refs.~\cite{DeRafael:1974iv,Broadhurst:1992za,Dunne:2002ta,Baikov:2012rr})
\beq
\beta(\ald) = \frac{\ald^2}{2\pi}\left[\frac{4}{3}\left(n_F + \frac{n_S}{4}\right) +
\frac{\ald}{\pi} (n_F + n_S)\right]\,,
\label{beta}
\eeq
where $\beta(\ald)\equiv \mu \, d\ald/d\mu$.  Here, $\mu$ is the renormalization
scale, which we will later take to be set by the momentum transfer
$q$ characterizing the interactions of $\gamd$.
\begin{figure}[t]
\includegraphics[width=0.46\textwidth,clip]{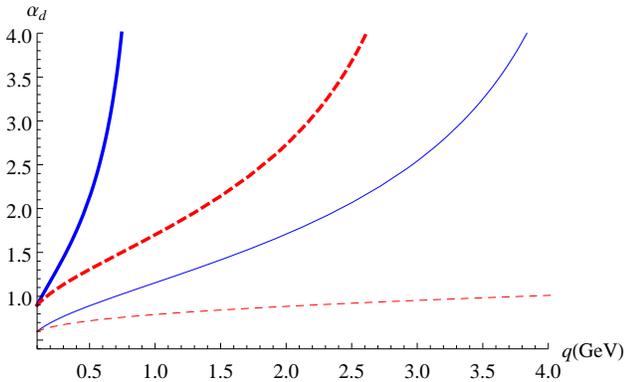}
\caption{Running of $\ald(q)$ in the basic model with one dark matter state and one dark Higgs boson.  The solid (dashed)
curves correspond to a fermion (scalar) dark matter state and the thin (thick) curves correspond to
$\ald(q_0) = 0.6$ (0.9), where $q_0=0.1$~GeV.  We have implicitly assumed $\mgd\lsim q_0$.}
\label{aldrun1DM}
\end{figure}

\begin{figure}[t]
\includegraphics[width=0.46\textwidth,clip]{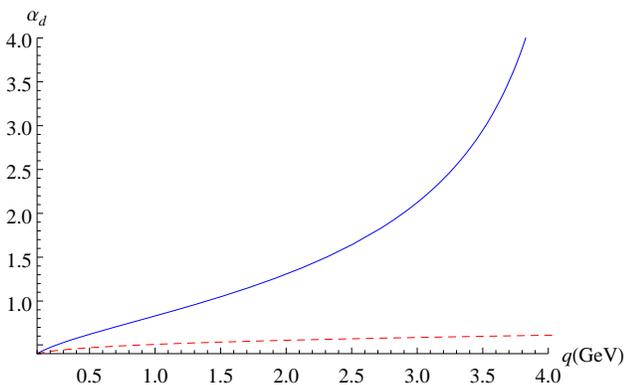}
\caption{Running of $\ald(q)$ with two dark matter states and one dark Higgs boson.  The solid (dashed)
curve corresponds to fermionic (scalar) dark matter and
$\ald(q_0) = 0.4$, where $q_0=0.1$~GeV, again assuming $\mgd\lsim q_0$.}
\label{aldrun2DM}
\end{figure}

In Fig.\ref{aldrun1DM}, we have plotted the running of $\ald$ in the basic model with one dark matter state.  Throughout
our analysis, a dark Higgs scalar $\Phi_d$, assumed to break $U(1)_d$, is included in
the running and hence $n_S\geq 1$ for all of our results.  We will consider a minimum
momentum transfer $q_0=100$~MeV, which is used to set the value of $\ald$
at the lower kinematic range of typical experiments.  The solid and dashed curves
correspond to fermionic and scalar dark matter [$(n_F, n_S)=(1,1)$ and $(0,2)$ in \eq{beta}], respectively;
$\ald(q_0)=0.6$ ($0.9$) is represented by the thin (thick) curve.  Based on the form of \eq{beta},
for $\ald\gsim \pi$ the perturbative analysis becomes unreliable, due to strong coupling.

As can be seen from the plot in Fig.\ref{aldrun1DM}, in all cases, except
for the one with $\ald(q_0)=0.6$ and $(n_F, n_S)=(0,2)$, the value of
$\ald$ grows large, {\it i.e.} $\ald \gsim \pi$, by the time
$q$ reaches $\sim$ a few GeV.  This is the typical kinematic domain of the proposed fixed target or beam dump experiments.
These results suggest that values of $\ald\gsim 0.6$, used to illustrate the phenomenology in some studies
\cite{Izaguirre:2013uxa,Diamond:2013oda,Izaguirre:2014bca},
can typically lead to unreliable predictions, unless those values correspond only to the upper kinematic range $q\gsim 1$~GeV.
In general, numerical predictions are more stable when only dark scalars are present in the low energy theory.
However, even in that case, corresponding to $(n_F,n_S)=(0,2)$, we see that
the change in $\ald$ is not negligible for $\ald(q_0)\gsim 0.6$.

In Fig.\ref{aldrun2DM}, we also present the results for running of $\ald$ in an extended model that has two
dark matter states.  See, for example, Ref.~\cite{Izaguirre:2013uxa} where such models have been
discussed as a viable setup for sub-GeV dark matter.  As can be seen from the figure, perturbative analysis becomes unreliable for
$\ald(q_0)\gsim 0.4$, with $q_0=100$~MeV, if dark matter states
are fermionic.  However, 2 scalar dark matter states do not
lead to a significant loss of perturbative validity.  Again, the running over $q \lsim$ few GeV is not
negligible, $\sim 50\%$, even for the all-scalar case and can
potentially have measurable effects.  Overall, we see that
the assumption of a constant $\ald$ over the kinematic range of typical experiments can lead
to underestimation of the predicted rates, even when $\ald$ is not close to unity.

The preceding results point to an interesting possibility at fixed target or beam dump experiments,
where a range of values for momentum transfer $q$ are accessible.  For
$\ald(q_0) \gsim 0.2-0.3$, measurements of processes at different $q$ can probe
the running of $\ald(q)$, in combination with the induced running of $\eps(q)$. As illustrated
above, the running of modest-sized $\ald$ is sensitive to the number and type - {\it i.e.} fermion versus scalar -
of dark states, as long as they are well below the typical scale of momentum transfer in the measured processes.

Definitive statements regarding the
measurement of the running with $q^2$ in the above scenarios
depend on the specifics, such as the production and detection processes,
and the energy spectrum of the light dark matter produced
by a beam dump or another intense source.  
Nonetheless, key general features of the dark matter scattering through
$\gamd$ exchange can be used to outline a potential
path toward such measurements, as explained below.  
Of course, much more detailed studies 
based on specific theoretical and  
experimental parameters would be 
warranted in designing a dedicated experiment, or 
upon discovery of light dark matter models considered here.  

Let us consider the on-shell production of $\gamd$, whose rate
is proportional to $\eps^2(\mgd)$, a constant set by $q^2=\mgd^2$.  Here,
$\gamd$ primarily decays into dark matter which then scatters in the detector with
a cross section $\sigma_{\rm DM} \propto \ald(q) \eps^2(q)$, where
$q$ has a distribution over some range.  In typical scenarios, kinetic mixing is loop-induced and $\eps^2(q) \propto \ald(q)$,
which gives $\sigma_{\rm DM}\propto \ald^2(q)$.

At $q \gsim \mgd$, dark matter interactions with
the nucleus are akin to electromagnetic interactions of a charged lepton
with the nucleus, governed by quantum electrodynamics (QED).
This correspondence can be used to obtain precise predictions for the
dark matter scattering cross section $\sigma_{\rm DM}$.  In particular, one could
normalize $\sigma_{\rm DM}$ to the electron (or muon) electromagnetic cross section $\sigma_{\rm EM} \propto 1/q^2$,
which is theoretically well-understood and can be precisely measured.
Thus, modulo QED radiative corrections and non-zero $\mgd$ propagator effects, we have (for $q\gsim \mgd$)
\beq
R\equiv \sigma_{\rm DM}/\sigma_{\rm EM} \simeq \ald\, \eps^2/\alpha
\simeq \xi\, \ald^2\, ,
\label{R}
\eeq
where $\xi$ is approximately a constant (we have ignored the negligible running of the QED coupling
$\alpha$ over the range of $q^2$ considered here).  With the above assumptions,
the value of $\xi$ can be known with good accuracy, given the input parameters $\eps$ and $\ald$ at a reference
value of momentum transfer, which we can naturally take to be $q=\mgd$.  In turn, for a given value of
$\xi$, one could predict the scattering cross section $\sigma_{\rm DM}$ and its dependence on $q^2$ quite well,
in our assumed framework.

\begin{figure}[t]
\includegraphics[width=0.46\textwidth,clip]{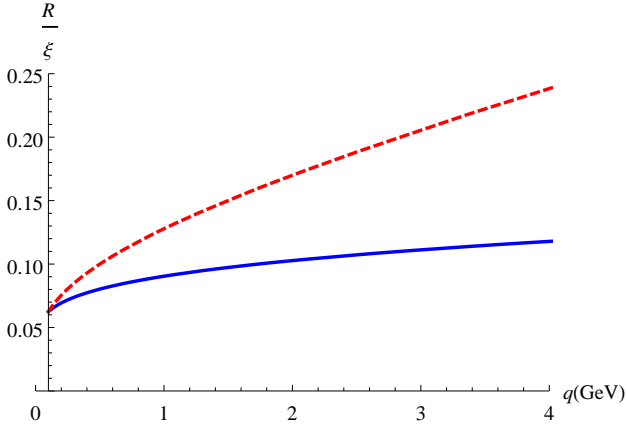}
\caption{
The running of $R/\xi$ as a function of momentum transfer $q$, assuming (A) one and (B) two
light fermionic dark matter states, corresponding to the solid and dashed curves, respectively.  We have set
$\ald(q_0) = 0.25$, $q_0=0.1$~GeV, and $\mgd\lsim q_0$ is assumed.
A scalar (dark Higgs boson) is included in the running for both cases.
}
\label{fig:R}
\end{figure}
The combined effect of $\ald(q)$ and $\eps^2(q)$ running $\propto \ald^2(q)$ can be significant.
For an illustrative numerical example, let us consider the case with $\ald(q_0) = 0.25$
(assuming $\mgd\lsim q_0$) and $q_0 = 100$~MeV.  With these parameters, the effect of running with $q^2$
on the event rate can be significant, depending on the
value of the $\beta$ function in \eq{beta}.  To show this, we consider the cases of (A) one dark fermion and (B) two
dark fermions contributing to the running (a dark Higgs scalar is included in each case).
We illustrate the running of $R/\xi = \ald^2$, for the two cases A and B above in Fig.\ref{fig:R}.
One can see that the running is significant, over the range $q \in [0.1, 4]$~GeV, in both cases: $R/\xi$ changes by
a factor of $\sim 2$ for case A (solid curve) and a factor of $\sim 4$ for case B (dashed curve).  Furthermore, 
if sufficient statistics are available 
one can easily distinguish between the two cases, as the figure shows, which can
potentially probe the dark sector spectrum over the measured $q^2$.

Here, we would like to add a comment.  The rise of the ratio $R$ with $q$ only
encodes the relative increase in $\sigma_{\rm DM}$
compared to the case with constant couplings.  However, $\sigma_{\rm DM}$ falls
like $1/q^2$, for $q\gsim \mgd$ and modulo $\ald$ running, and hence
it is expected that the dark matter scattering signal would be stronger for
lower values of $q^2$, whereas 
potential backgrounds from neutrino-nucleus scattering become more
suppressed.  The optimal range of $q^2$ for detecting the
dark matter scattering signals depends on the details of the experimental
setup.  However, as long as that range is moderately broad, the $q^2$ running
could in principle be measurable, as implied by the plot in Fig.\ref{fig:R}.

So far, we have limited our discussion to the running of $\ald$ over the GeV-scale values of $q$,
relevant to predictions for proposed fixed target (beam dump) experiments.  In practice, one
may not worry if $\ald$ becomes too large and approaches a Landau pole at $q^*\gg 1$~GeV, as far as
those predictions are concerned.  However, we will argue below that non-perturbative values of $\ald$
should typically be postponed to $q^*$ above the weak scale.
To see this, note that a straightforward way to resolve the problem of an ever-growing $\ald$ is
to assume the appearance of a new non-Abelian gauge interaction $G_d$ at $q > q^*$,
whose breaking yields $U(1)_d$ at lower energies.  In typical scenarios, the kinetic mixing
parameter $\eps$ vanishes at $q^*$, as required by gauge invariance \cite{dim5}.  However, to observe any
events in fixed target experiments, often $\eps \gsim 10^{-4}$ is required.
This implies that $\eps$ must run to non-zero values {\it below} $q^*$, due to the effects
of new states, denoted by $F_{d,Y}$ in Fig.\ref{eps-run},
which generate $\eps\neq 0$ at the quantum level \cite{Holdom:1985ag}.  These new states
must be charged under both $U(1)_d$ and hypercharge $U(1)_Y$, which means that they cannot be lighter
than $\sim 100$~GeV, or else they would have been discovered in high energy experiments \cite{Davoudiasl:2012ig}.

In light of the above theoretical consistency conditions for $\eps\neq 0$ at low energies, we
find it well-motivated to require that $\ald$ should remain perturbative up to $q^*\gsim 100$~GeV.  This requirement
implies an upper bound on the low energy value of the $U(1)_d$ coupling $\ald (q_0)$.  One can derive an estimate
of this upper bound, using the 1-loop running equation
\beq
\ald (q_0) = \frac{\ald (q^*)}{1 + \frac{2}{3\pi}\ald (q^*)(n_F + n_S/4)\ln(q^*/q_0)}.
\label{aldq}
\eeq
The value of $\ald (q_0)$ becomes insensitive to the high scale value $\ald (q^*)\gsim 1$, the onset
of strong coupling, as long as $\ln(q^*/q_0)\gg 1$, and we get
\beq
\ald (q_0) \approx \frac{3\pi}{(2 n_F + n_S/2)\ln(q^*/q_0)}.
\label{aldqest}
\eeq
In the above, we have implicitly assumed that $\mgd \lsim q_0$ and we will consider, as before,
that $q_0=0.1$~GeV, a typical value in the lower kinematic range for fixed target experiments.

We see that for $q^*=100$~GeV, as a minimal requirement based on our preceding discussion,
an upper bound $\ald(q_0) \lsim 0.68/(n_F + n_S/4)$ is obtained.  Hence, for moderate
values of $n_F$ and $n_S$ one can obtain interesting upper bounds.  For instance,
if $n_F = 1$ and $n_S=1$ (corresponding to a dark Higgs), we find $\ald(q_0) \lsim 0.5$; this upper bound 
gets reduced to $\ald(q_0)\lsim 0.3$ for $n_F=2$.  One may entertain
much larger values of $q^*$, potentially near the Planck scale $\mP \approx 1.2 \times 10^{19}$~GeV.  This
could be motivated if one assumes that there is no new physical mass scale above the electroweak scale,
which may possibly address the stability of the SM Higgs mass against large quantum corrections
\cite{Bardeen:1995kv,Farina:2013mla}.  In that case,
one gets $\ald(q_0) \lsim 0.1/(n_F + n_S/4)$ which
could place stringent upper bounds on $\ald(q_0)$, in the scenarios considered in this work.  In particular,
a dark matter interpretation of the $\gamd$ invisible final state implies
the relation among model parameters given in \eq{ald}.  For $\eps \lsim 10^{-4}$,
accessible to future experiments, that relation would typically require $\ald\gsim 0.1$, which would be
in perturbative tension with $q^*\sim \mP$.

In passing, we observe that based on constraints
from Cosmic Microwave Background Radiation (CMBR) measurements \cite{Madhavacheril:2013cna},
$p$-wave annihilation of GeV-scale thermal relic dark matter is favored, since
it becomes less efficient at smaller velocities characteristic of the CMBR decoupling era \cite{Batell:2014mga}.
This requirement points to $U(1)_d$-charged light scalar states as good thermal relic dark matter
candidates \cite{Boehm:2003hm,deNiverville:2011it}.
However, excess annihilation that could distort CMBR could also be avoided if
dark matter density is given by an asymmetry at late times, which would preclude the possibility
of conjugate-pair annihilation.  We would also like to add that in the models considered in our work,
dark matter scattering from itself or other dark states, mediated by $\gamd$, becomes stronger at earlier epochs
with higher temperatures and larger characteristic values of $q$.  In general,
the running of dark sector couplings could lead to potentially
interesting effects in early universe cosmology \cite{Zhang:2009dd,Sannino:2014lxa},
but we will not further speculate on this question here.

To summarize, we have considered the running of a dark sector $U(1)_d$ fine structure constant
$\ald$ due to the presence of light dark particles in quantum loop corrections.
Invisible dark photons, assumed to decay on-shell to dark matter states,
belong to this class of models.  We observed that the running of $\ald(q)$
can lead to a running kinetic mixing parameter $\eps^2(q)$,
for it is naturally loop-induced and therefore proportional to $\ald(q)$.
Under some minimal well-motivated assumptions about the
dark sector content, we find that, roughly speaking, values of $\ald\gsim 0.4$ can lead to
departure from a perturbative analysis and unreliable predictions over the
range of momentum transfer 10~MeV$\lsim q \lsim 1$~GeV
in proposed fixed target or beam dump experiments.
We note that $\ald$ values not much smaller than $\sim 1$ are typical in scenarios 
with sub-GeV dark matter particles that are 
lighter than the dark photon.  Light fermionic dark states lead to faster running of $\ald$ compared
to scalar states and can result in loss of perturbative control over larger regions of
parameter space.

We pointed out that the dependence of $\ald$ on $q$ can be used
to help probe the low-lying spectrum of the dark sector in fixed target or beam dump
experiments.  This can be done if measurement of the event rates as a function of $q$ is feasible, over a moderately
broad range of $q^2$. Those rates probe the combined running of $\ald(q)$
and $\eps^2(q)\propto \ald(q)$ with momentum transfer,
leading to a potentially significant sensitivity $\propto \ald^2(q)$.  We showed that for infrared
values $\ald(q_0) \gsim 0.2$, one could expect significant effects on the event rate from the running, in typical
scenarios.  A thorough discussion of $q^2$ running measurements would require 
input from specific experimental parameters and would certainly be warranted 
upon the discovery of a light dark matter signal from an intense source.  
However, on general grounds, we discussed how electromagnetic cross sections
for electron (or muon) scattering from the nucleus can be used to 
obtain precise predictions for the corresponding dark matter
scattering cross section, in the kinetic mixing scenario considered.  

We also argued that theoretic considerations imply the perturbative range for $\ald$ should extend
to values of $q$ at or above the weak scale, perhaps even the Planck scale, in which case tighter upper bounds
on the low energy value of $\ald$ and the dark sector spectrum can be obtained.

\begin{acknowledgments}

Work supported by the US Department of Energy under Grant Contract DE-SC0012704.
\end{acknowledgments}

\end{document}